\documentclass[prl,twocolumn]{revtex4}
\usepackage{graphicx}
\usepackage{dcolumn}
\usepackage{bm}
\usepackage{color}
\usepackage{eso-pic}
\usepackage{amsmath}
\usepackage{xspace}
\newcommand{\STO}{$\mathrm{SrTiO}_3$\xspace}
\newcommand{\LAO}{$\mathrm{LaAlO}_3$\xspace}
\newcommand{\LTO}{$\mathrm{LaTiO}_3$\xspace}
\newcommand{\LCO}{$\mathrm{LaCrO}_3$\xspace}
\newcommand{\LCxO}{$\mathrm{LaAl}_{1-x}\mathrm{Cr}_x\mathrm{O}_3$\xspace}

\usepackage{soul}

\begin{document}
\title{Effect of disorder on superconductivity and Rashba spin-orbit coupling in \LAO/\STO interfaces.}
\author{G. Singh$^{1,2}$, A. Jouan$^{1,2}$, S. Hurand$^{1,2}$, C. Palma$^{1,2}$, P. Kumar$^3$, A. Dogra$^3$, R. Budhani$^{3,4}$, J. Lesueur$^{1,2}$, N. Bergeal$^{1,2}$}
\affiliation{$^1$Laboratoire de Physique et d'Etude des Mat\'eriaux, ESPCI Paris, PSL Research University, CNRS, 10 Rue Vauquelin - 75005 Paris, France.\\
$^2$Universit\'e Pierre and Marie Curie, Sorbonne-Universit\'es,75005 Paris, France.
$^3$National Physical Laboratory, Council of Scientific and Industrial Research (CSIR)
Dr. K.S. Krishnan Marg, New Delhi-110012, India.\\
$^4$Condensed Matter Low Dimensional Systems Laboratory, Department of Physics, Indian Institute of Technology, Kanpur 208016, India.}
\begin{abstract}
A rather unique feature of the two-dimensional electron gas (2-DEG) formed at the interface between the two insulators  \LAO and \STO is to  host both gate-tunable superconductivity and strong spin-orbit coupling. In the present work, we use the disorder generated by Cr substitution of Al atoms in \LAO as a tool to explore the nature of superconductivity and spin-orbit coupling in these interfaces. A reduction of the superconducting $T_c$ is observed with Cr doping consistent with an increase of electron-electron interaction in presence of disorder. In addition, the evolution of spin-orbit coupling with gate voltage and Cr doping suggests a  D'Yakonov-Perel mechanism of spin relaxation in the presence of a Rashba-type spin-orbit interaction.
\end{abstract}

\maketitle

Oxide interfaces involving band or Mott insulators such as \LAO/\STO and \LTO/\STO have attracted much attention in recent years due to the formation of a two-dimensional electron gas (2-DEG) \cite{ohtomo} whose quantum properties include in particular, superconductivity \cite{reyren, biscaras}, strong spin-orbit coupling  (SOC) \cite{cavigliaSO, shalom} and magnetism \cite{li,bert}. The interplay between these different properties make \STO based interfaces particularly interesting from  fundamental as well as technological perspectives \cite{hwang,takagi,mannhart,bibes}. The conducting interface is identified as two-dimensional in nature, with a typical thickness $\sim$10 nm, smaller than the Fermi wavelength in the normal state, and smaller than the superconducting coherence length \cite{basletic,reyren,biscaras}. Electrons are confined within a quantum well which accommodates different subbands built on the anisotropic $t_{2g}$ orbitals of the Ti ions. A key advantage of these interfacial 2-DEG lies in the possibility to tune continuously  the electron density by electric field effect \cite{thiel}.  As a result, gate-tunable superconductivity was achieved both in back-gate \cite{caviglia,biscaras2,thiel} and top-gate geometry \cite{hurand, jouan} and, for strong carrier depletion, a superconductor-to-insulator quantum phase transition was observed \cite{caviglia,SIT}. Transport measurements indicate that the nature of the SOC is most likely of the Rashba type as expected in an asymmetric quantum well \cite{cavigliaSO,shalom,hurand}. The SOC strength reaches values of several meV, which is significantly larger than the ones found in semiconducting heterostructures. In addition, its intensity can be modulated with a gate voltage.\\
\indent Because of 2D confinement, it is expected that the electronics properties of these interfaces should be strongly affected by external dopants, hence providing a way to explore their fundamental nature.  Fix et al. reported a quenching of electron density and mobility in \LAO/SrTi$_{1-x}$R$_x$O$_3$ when doped by Mn \cite{fix,schoofs,gray} while, Sanders et al. recently observed that 2$\%$ doping of rare earth ions Tm and Lu at the La site of \LAO does not significantly affect the electronic transport \cite{sanders}. An increase of SOC in $\delta$ doped  \LTO/\LCO/\STO interfaces was reported by Das et al. \cite{das}. 
In these reports, it appears difficult to disentangle the respective roles of doping, disorder and carrier density changes on the transport properties of the 2-DEG.

In this article, we present a detailed analysis of gate dependence of superconductivity and SOC in \LAO/\STO interfaces by controlled site substitution of Al by Cr.  \LCO is an antiferromagnetic band insulator with a Neel temperature of 290 K and the \LCO/\STO heterostructure is  insulating. However, the \LCxO/\STO interface was found to be conducting for the doping level used in this study \cite{chambers,colby}. We first show that Cr doping essentially increases elastic scattering, that is the atomic scale disorder, without changing the carrier density. As a consequence, superconductivity is weakened, and disappears according to Finkelstein's theory \cite{finkelstein} for which disorder is the only relevant parameter. On the other hand, we confirmed through magnetoconductance that spin relaxation occurs by mean of a D'ykonov-Perel mechanism \cite{perel}, and that the spin diffusion length is independent of the disorder as expected in that case.

 For this study, three 10 u.c thick \LCxO  films with doping  $x$ =0, 0.1 and 0.2 were grown on $\mathrm{TiO}_2$ terminated (001) \STO substrate. The thin films were deposited at a constant substrate temperature of 750 $^0$C in an oxygen pressure of 10$^{-4}$ mbar. During the deposition, the layer by layer growth was monitored by high energy electron diffraction (RHEED). Further details about the growth and structural characterization can be found in previous reports \cite{kumar}. A metallic gate is deposited on  the backside of  each \STO substrate, and standard four probe transport measurements were performed in a dilution refrigerator with a base temperature of 15 mK, equipped with a 6 T magnet. 
\begin{figure}[t]
\begin{center}
\vskip -0.8cm
\includegraphics [width=8.5cm]{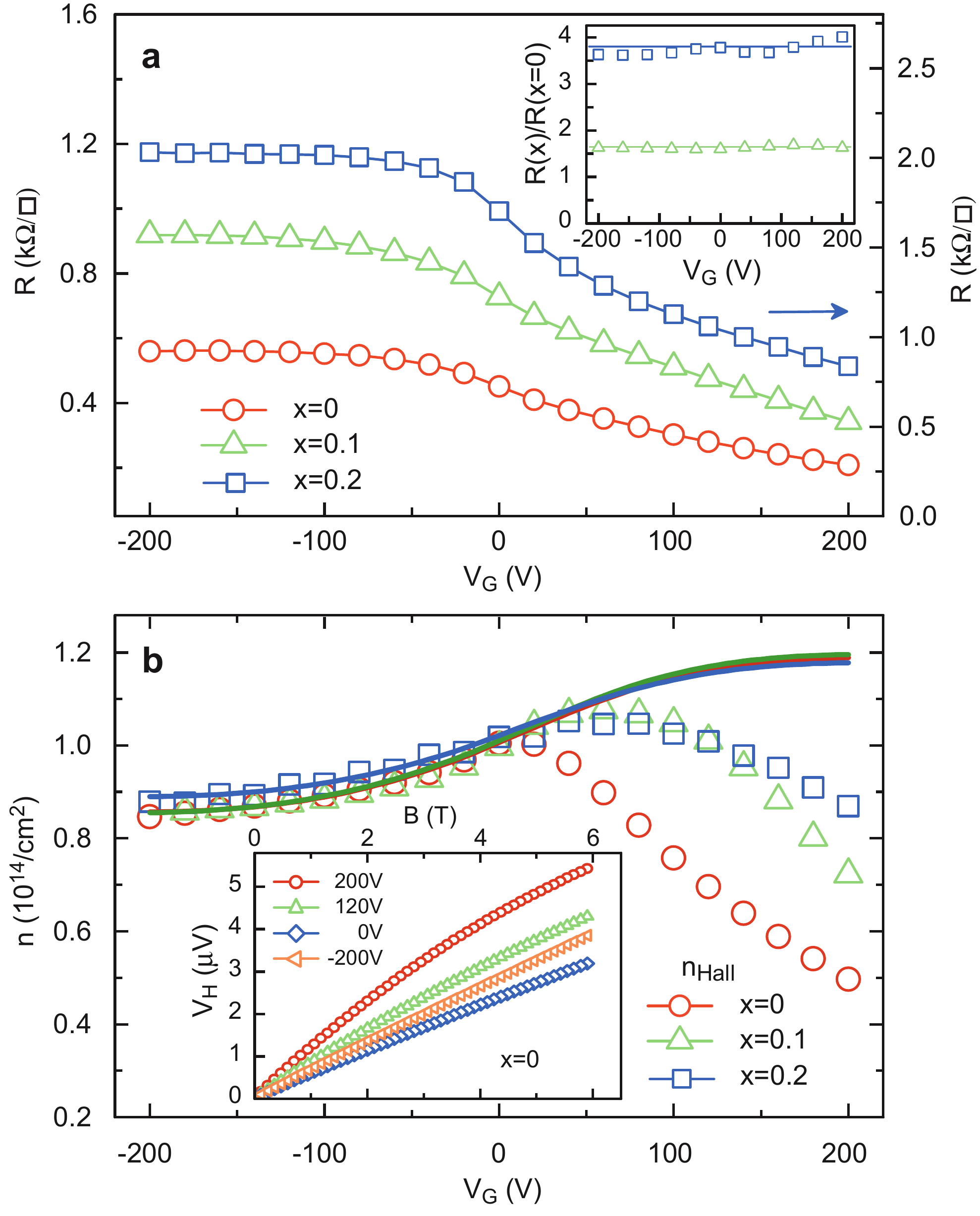}%
\end{center}
\vskip -0.5cm
\caption{(a) Sheet resistance $R$ measured at $T$ = 3 K as a function of gate voltage $V_\mathrm{G}$ for  three \LCxO/\STO samples corresponding to doping $x$=0 (left axis), $x$=0.1 (left axis) and $x$=0.2 (right axis). Inset) Sheet resistance of the $x=0.1$ and $x=0.2$ samples normalized by the sheet resistance of the $x=0$ sample. Straight lines are guides for the eyes.  (b) Carrier density extracted at $T$ = 3 K from low-magnetic-field Hall effect measurements ($n_\mathrm{H}$, open symbols) and from capacitance measurements ($n$, full lines, same color code)\cite{SM}. Inset) Hall voltages V$_\mathrm{H}$ as a function of the magnetic field $B$ of the $x$=0 sample measured for different $V_\mathrm{G}$.} 
\end{figure}

We first investigate the effect of Cr doping on the sheet resistance and carrier density of the \LCxO/\STO interfaces. 
 After cooling the sample to 3K, the back-gate voltage was first swept to its maximum value +200V while keeping the 2-DEG at the electrical ground, to insure that no hysteresis will take place upon further gating \cite{biscaras3}. As expected when depleting the quantum well,  the sheet resistance $R$ of the three samples continuously rises when the gate voltage is decreased down to -200V (Fig 1a). The absolute value of the sheet resistance for a given gate voltage $V_\mathrm{G}$ increases with $x$, indicating either an enhancement of scattering or a reduction of the carrier density with Cr doping.

 \indent To determine the carrier density in the 2-DEG, we performed Hall effect measurements as a function of $V_\mathrm{G}$ for the different samples. As already reported in the literature \cite{biscaras2,kim,ohtsuka}, the Hall voltage is linear in magnetic field in the underdoped regime  ($V_\mathrm{G}$ $<$  0) (inset Fig. 1b), and the carrier density is correctly extracted from the slope of the Hall voltage V$_\mathrm{H}$  (i.e. n$_\mathrm{H}$= IB/eV$_\mathrm{H}$ where $I$ is the bias current  and $B$ the magnetic field). This is no longer the case in the overdoped regime ($V_\mathrm{G}>0$), where V$_\mathrm{H}$ is not linear with $B$, sign of a multibands transport. In this case n$_\mathrm{H}$ measured at low magnetic field (see symbols in Fig 1b) is not the carrier density. This explains why n$_\mathrm{H}$ surprisingly decreases at high gate voltage, while more electrons are added in the quantum well. The true variation of the carrier density $n$ can be retrieved by measuring the capacitance $C$ between the 2-DEG and the gate\cite{biscaras2,SM}. The added charges correspond to the integral of $C$ over the gate voltage range (solid lines in Fig. 1b). The absolute value $n$ is calculated by assuming that for negative voltages (one band transport ) $n$=n$_\mathrm{H}$, which is well verified experimentally  (Fig. 1b). Note that $n\sim1\times$~$10^{14}$  $e^{-}\cdot\mbox{cm}^{-2}$ for $V_\mathrm{G}=0$, a value not too far from the number predicted by the polarization catastrophe scenario ($\simeq$ 3.3~$\times$~$10^{14}$ $e^{-}\cdot\mbox{cm}^{-2}$) \cite{nakagawa}. 

 Remarkably, as evidenced in Fig. 1b, the sheet carrier density \emph{n does not depend on Cr doping} within the error margins in this doping range (higher Cr doping leads to a significant reduction  of carrier density \cite{kumar}). The increase of  sheet resistance R with $x$ is therefore mainly due to a reduction of the elastic scattering time $\tau_e$. Within a simple Drude model, one expects $R(x)/R(x=0)$=$\tau_{e}($x$=0)/\tau_{e}(x)$ to be constant, independent of the gate voltage. This is verified in inset of Fig. 1a, where $R(x)/R(x=0)$ is plotted for $x$=0.1 and $x$=0.2. Small deviations occur for $V_\mathrm{G} > 0$ due to multiband transport. The first conclusion of this study is that \emph{Cr doping increases the atomic disorder in the system}, and, at first order, only modifies the elastic scattering time $\tau_e$. In the following, we will study its influence on superconductivity and SOC.

The $x$=0 sample displays a gate-tunable superconducting transition  (Fig. 2a) as usually observed in \STO based interfaces. A superconducting phase diagram can be drawn by plotting the transition temperature $T_c$ (mid-point) extracted from R(T) curves, as a function of $V_\mathrm{G}$. It exhibits a partial dome shape with a maximum $T_c$ of approximately 170 mK at an optimal electrostatic doping of $V_\mathrm{G}\sim$80 V (Fig 2c).  For the heavily doped sample ($x$=0.2), no sign of superconductivity can be seen down to 20 mK  (not shown).  At intermediate doping ($x$=0.1), the sample shows broad resistive transitions always saturating to a residual resistance at low temperature (Fig. 2b). 
Despite signs of a superconducting condensation (such as the typical lowering of  $R$ below $T_c$), superconductivity is too weak to establish phase coherence across the whole sample. Nevertheless, a phase diagram similar to the one of the $x$=0 sample is obtained, with a reduced mid-point $T_c$ (Fig. 2c). In the underdoped regime (i.e. for gate values lower than the one maximizing $T_c$), incomplete resistive transitions in \STO based interfaces are systematically reported in the literature.
 
To explain this behavior, Caprara \textit{et al.} suggested that these electronic systems should rather be described as an array of superconducting islands with a random distribution of $T_c$ coupled through a metallic 2-DEG by proximity effect \cite{caprara,scopigno,biscaras4,SIT}.  At sufficiently low electrostatic doping, the superconducting fraction of the system becomes too low to enable a percolative superconducting transition, and, despite a sizable drop in resistance, a true zero resistive state is never reached at any arbitrary low temperature. Such scenario is  illustrated by the $R(T)$ curves  measured on the $x$=0 sample at negative gate voltages (Fig. 2a). 

The disorder induced by Cr doping $x$, i.e. the decrease of the elastic scattering time $\tau_e$, will affect both the superconducting islands and the metallic 2-DEG.  On one hand, it lowers the Josephson coupling between islands : in diffusive proximity effect, the normal coherence length reads $\xi_n=\sqrt{\frac{\hbar D}{k_BT}}$ where the electronic diffusion constant $D$ is proportional to $\tau_e$. This reduced coupling explains why the resistive transitions are broader in Cr doped samples, and why the R=0 state is never reached. On the other hand, it weakens superconductivity inside the islands, which is essentially measured by the maximum $T_c$. According to Finkelstein\textquoteright s theory \cite{finkelstein}, disorder and electron-electron interactions lower $T_c$ following a universal law that depends only on the sheet resistance (inversely proportional to $\tau_e$). Fig. 2d shows $T_c$ around the optimal electrostatic doping as a function of $R$ for the different samples, together with the Finkelstein's function (see Supplementary Material for the fitting procedure \cite{SM}). A quantitative agreement is obtained, with a $T_c$ = 350 mK in the limit of null disorder, close to the one of doped bulk \STO. It indicates that the enhanced disorder through Cr doping can explain the disappearance of superconductivity in doped samples.\\


\begin{figure}[t]
\begin{center}
\vskip -0.2cm
\includegraphics [width=8.5cm]{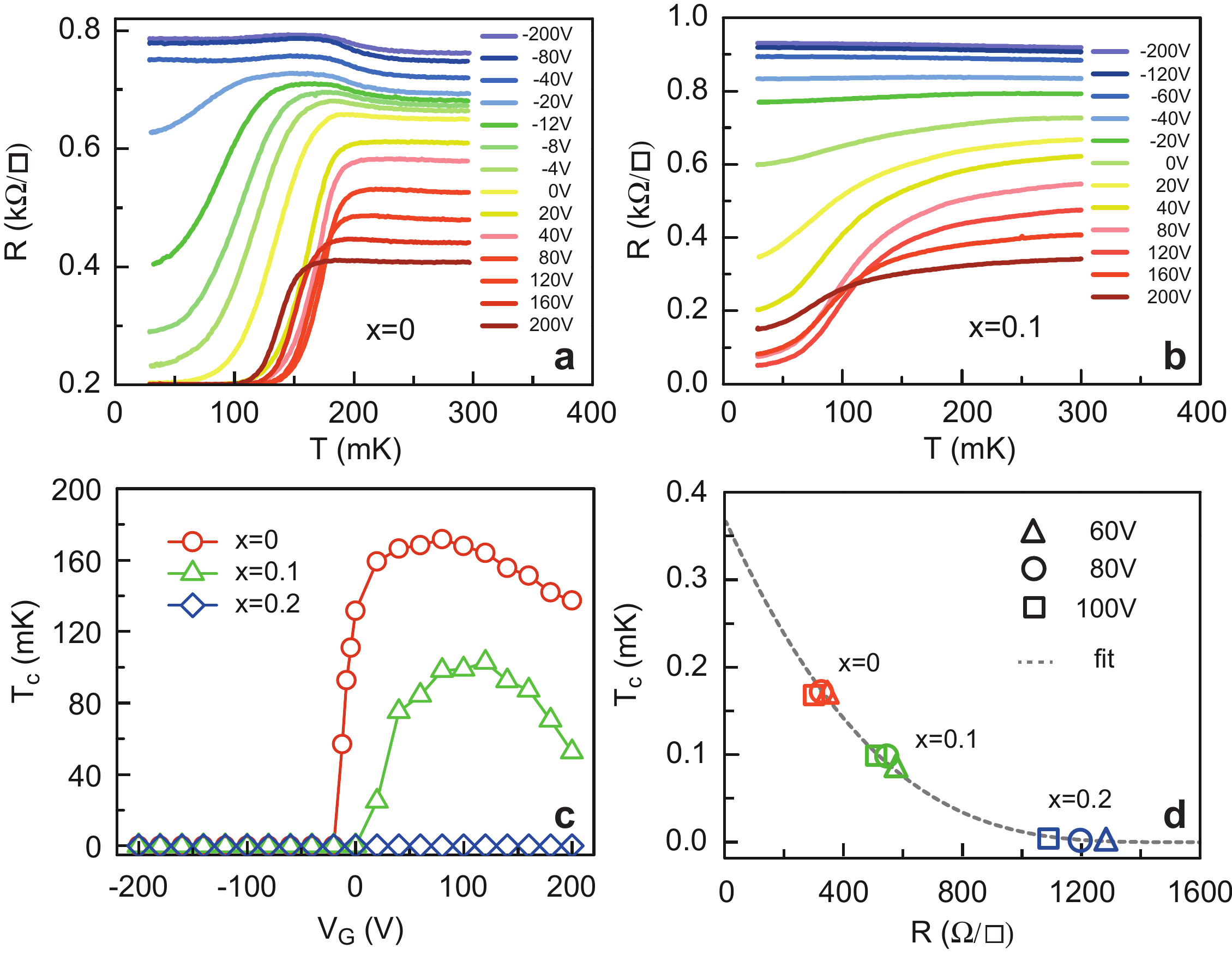}%
\end{center}
\vskip -0.5cm
\caption{(a) Sheet resistance $R$ vs temperature $T$ of the $x$=0 sample for different $V_\mathrm{G}$. (b) Same for the $x$=0.1 \LCxO/\STO  sample. c) Superconducting  $T_c$ (defined as a 50$\%$ drop of R) as a function of $V_\mathrm{G}$ for the three samples. d) $T_c$ as a function of the sheet resistance for the three samples and for three different $V_\mathrm{G}$ corresponding to the optimal electrostatic doping (i.e. the top of the T$_c$ dome). Data are fitted by the Finkelstein\textquoteright s theory (see Supplementary Material \cite{SM} (dotted line). }
\end{figure}

\begin{figure}[t]
\begin{center}
\vskip 0.5cm
\includegraphics [width=8.5cm]{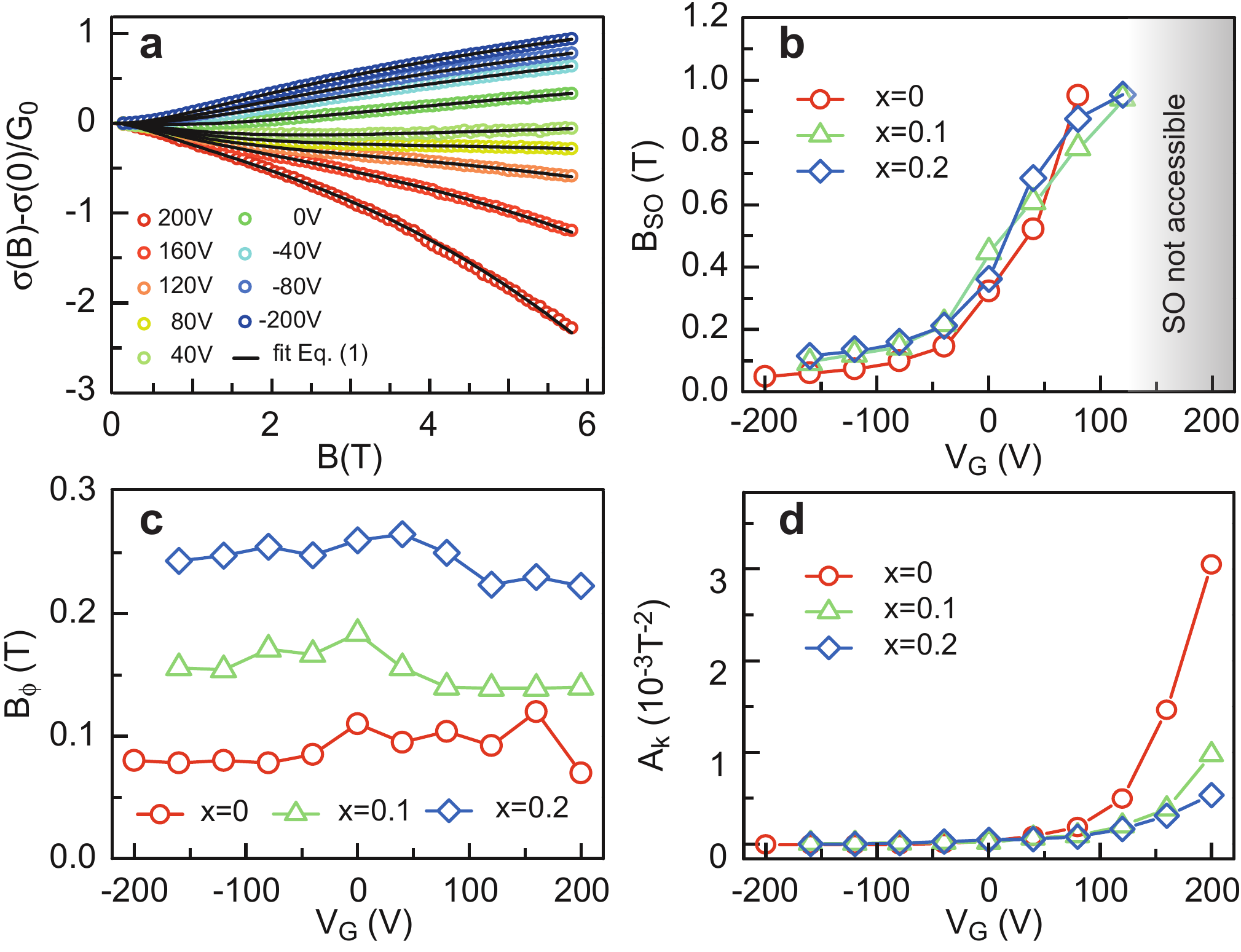}%
\end{center}
\vskip -0.7cm
\caption{\label{fig2}  (a) Normalized magneto-conductance as a function of out-of-plane magnetic field $B$ of the $x$=0.2 sample for different $V_\mathrm{G}$ (open symbols) fitted with Eq. (1). (b,c,d)  Fitting parameters $B_\mathrm{SO}$, $B_\phi$ and $A_K$  as a function of $V_\mathrm{G}$ for the three samples ($x$= 0, 0.1, 0.2). In the grey zone in Fig. 3b, B$_\mathrm{SO}$ and B$_\phi$ are determined with reduced accuracy (see text).}
\end{figure}

The confinement of electrons at the  \LAO/\STO interface generates a strong local electric field $E_z$ perpendicular to the motion of the electrons, which   gives rise to a Rashba-type spin-orbit coupling. It is described by the Hamiltonian $H_\mathrm{R}=\alpha(k_y\sigma_x-k_x\sigma_y)=\vec{B}_R(\vec{k})\cdot\vec{\sigma}$, where $\alpha$ is the coupling constant, $\vec{\sigma}$ are the Pauli matrices and  $\vec{B}_R(\vec{k})$ is the Rashba magnetic field whose direction and amplitude depend on the electron momentum $\vec{k}$ \cite{bychkov,winkler}. The spin of an electron propagating in the $E_z$ electric field precesses around $\vec{B}_R(\vec{k})$ between two scattering events which causes a continuous spin dephasing, a mechanism known as DÕyakonov-PerelÕ relaxation process \cite{perel}.  The randomization of the precession direction by collisions results in a corresponding spin relaxation time $\tau_\mathrm{SO}$ that varies inversely  with the elastic scattering time $\tau_e$. This is different from the Elliot-Yafet mechanism where the collisions of an electron with an impurity can generate a direct spin flip which is characterized by a relaxation time $\tau_\mathrm{SO}$ proportional to $\tau_e$.\\
\indent In a 2D system, $\tau_\mathrm{SO}$ can be evaluated by analyzing  how this additional spin relaxation mechanism modifies the weak localization corrections to the magnetoconductance $\Delta\sigma(B)=\sigma(B)-\sigma(0)$  \cite{maekawa,hikami}. In the following, we analyze the effect of Cr doping on SOC by measuring the conductance of the 2-DEG as a function of a magnetic field applied perpendicular to sample plane. The curves were fitted with Maekawa-Fukuyama formula \cite{maekawa,hurand} in the diffusive regime, neglecting the Zeeman splitting :
\begin{eqnarray}
\lefteqn{\frac{\Delta \sigma (B)}{G_{0}}=}\nonumber\\
& &-\psi \left [ \frac{1}{2}+\frac{Be}{B} \right ]+\frac{3}{2}\psi \left [ \frac{1}{2}+\frac{B_{\phi }+B_{SO }}{B} \right ]\nonumber\\
& & -\frac{1}{2}\psi\left [ \frac{1}{2}-\frac{B_{\phi }}{B} \right ]-\left [ \frac{B_{\phi }+B_{SO }}{Be} \right ]\nonumber\\
& & -\frac{1}{2}ln\left [ \frac{B_{\phi }+B_{SO }}{B_{\phi }} \right ] -A_{K}\frac{\sigma (0)}{G_{0}}B^{2}
\end{eqnarray}

Here, $\psi$ is the digamma function, $G_{0}=\frac{e^{2}}{\pi h}$ is the quantum of conductance and, B$_{e}$, B$_\phi$, B$_\mathrm{SO}$ are the elastic, inelastic and spin-orbit effective fields respectively. They are related to the elastic scattering time $\tau_e$, the inelastic scattering time $\tau_\Phi$ and the spin-orbit relaxation time $\tau_\mathrm{SO}$ by the expressions $B_e=\hbar/(4eD\tau_e)$, $B_\Phi=\hbar/(4eD\tau_\Phi)$ and $B_\mathrm{SO}=\hbar/(4eD\tau_\mathrm{SO})$ where $D=\frac{1}{2}v_F^2\tau_e$ is the electronic diffusion constant in two dimensions ($v_F$ is the Fermi velocity). A$_K$ is the Kohler term  that accounts for orbital magnetoconductance. Fig. 3a displays the magnetoconductance of the $x$=0.2 sample measured at T= 3K, for $V_\mathrm{G}$ ranging between $\pm$200V, and fitted by Eq. (1). A very good agreement is obtained between the experimental data and the theory over the whole electrostatic gating range for this sample as well as for the other ones (see Supplementary Material).\\
\indent For large negative $V_\mathrm{G}$, a positive magnetoconductance is observed consistently with  weak localization in the presence of a weak SOC. When $V_\mathrm{G}$ is increased, the magnetoconductance becomes negative because of enhanced SOC. Beyond $V_\mathrm{G}\sim$150V, the mobility of the 2-DEG increases, and so does the A$_K$ coefficient extracted from the fit (see Fig. 3d), which is proportional to the square of it\cite{macdonald}. In that situation, Kohler's contribution dominates over the weak localization correction, and the determination of B$_\phi$ and B$_\mathrm{SO}$ becomes less accurate.

Fig. 3c shows the inelastic field $B_\phi$ as a function of gate voltage for different Cr doping $x$=0, 0.1 and 0.2. $B_\phi$ does not show any significant dependence on $V_\mathrm{G}$ for the three samples, but increases clearly with doping $x$. This is a direct consequence of the enhancement of disorder, and therefore the reduction of the elastic scattering time $\tau_e$. Indeed, $B_\phi$ is inversely proportional to $\tau_e\times\tau_\Phi$ ($\tau_e$ is included into the diffusion constant $D$).  The inelastic scattering time $\tau_\Phi$ is independent of $x$ since it is dominated by electron-electron interactions \cite{hurand,biscaras}. In this case,  we expect that the variation of B$_\phi$ with $x$ is mainly contained into the variation of $\tau_e$. This is rather well verified within the experimental margin errors (see Supplementary Figure 2 \cite{SM}). In the framework of weak localization, $B_\phi$ corresponds to an inelastic scattering length $\ell_\phi=\sqrt{D\tau_\phi}=\large[\hbar/(4eB_{\phi})\large]^{1/2}$, which is plotted in Fig. 4a for different doping $x$. It is rather constant as a function of gate voltage, and estimated to be  $\ell_\phi\simeq$ 45 nm for $x$=0, $\ell_\phi\simeq$ 32 nm for $x$=0.1 and $\ell_\phi\simeq$ 25nm for $x$=0.2. Interestingly, when $\ell_\phi$ becomes smaller than the estimated bare superconducting coherence length $\xi\simeq$ 35 nm \cite{biscaras}, superconductivity is strongly suppressed and $T_c$ goes to zero (for $x$=0.2). This is coherent with the Finkelstein's scenario where enhanced electron-electron interactions due to disorder destroy superconductivity \cite{finkelstein,lee}.\\

The spin-orbit field $B_\mathrm{SO}$ increases monotonically with $V_\mathrm{G}$ and ranges between 0.1 to 1 T (Fig. 3b) as already observed in undoped samples \cite{hurand,herranz,cavigliaSO}. It is remarkable that despite the strong differences discussed above in the transport properties ($R$, $T_c$ and $B_\phi$ for instance), the  $B_\mathrm{SO}$ term is quasi-identical for the three samples in the whole range of gating.  This confirms experimentally that the D'Yakonov-Perel mechanism of spin relaxation \cite{perel} in the presence of Rashba spin-orbit interaction is dominant in these 2-DEG \cite{bychkov}. In this case, the reduction of the diffusion constant ($D\propto\tau_e$) because of disorder is exactly compensated by the increase of the spin-relaxation ($\tau_\mathrm{SO}\propto\frac{1}{\tau_e}$). As a result,  B$_\mathrm{SO}$ field (Fig. 3b) and the corresponding spin diffusion length $\ell_\mathrm{SO}=\sqrt{D\tau_\mathrm{SO}}$ (Fig. 4a) are essentially independent of disorder. For the lowest electron density $\ell_\mathrm{SO}=\large[\hbar/(4eB_\mathrm{SO}\large]^{1/2}\simeq$ 40-50 nm is comparable to the inelastic scattering length $\ell$$_\phi\simeq$ 25-40 nm (Fig. 4a). When the 2-DEG is strongly electrostatically doped, $\ell_\mathrm{SO}$ becomes smaller than $\ell_\phi$ ($\ell_\mathrm{SO}\simeq$ 15nm), indicating that the spin relaxation due to spin-orbit coupling dominates the decoherence  processes.

\begin{figure}[t]
\begin{center}
\includegraphics [width=8.5cm]{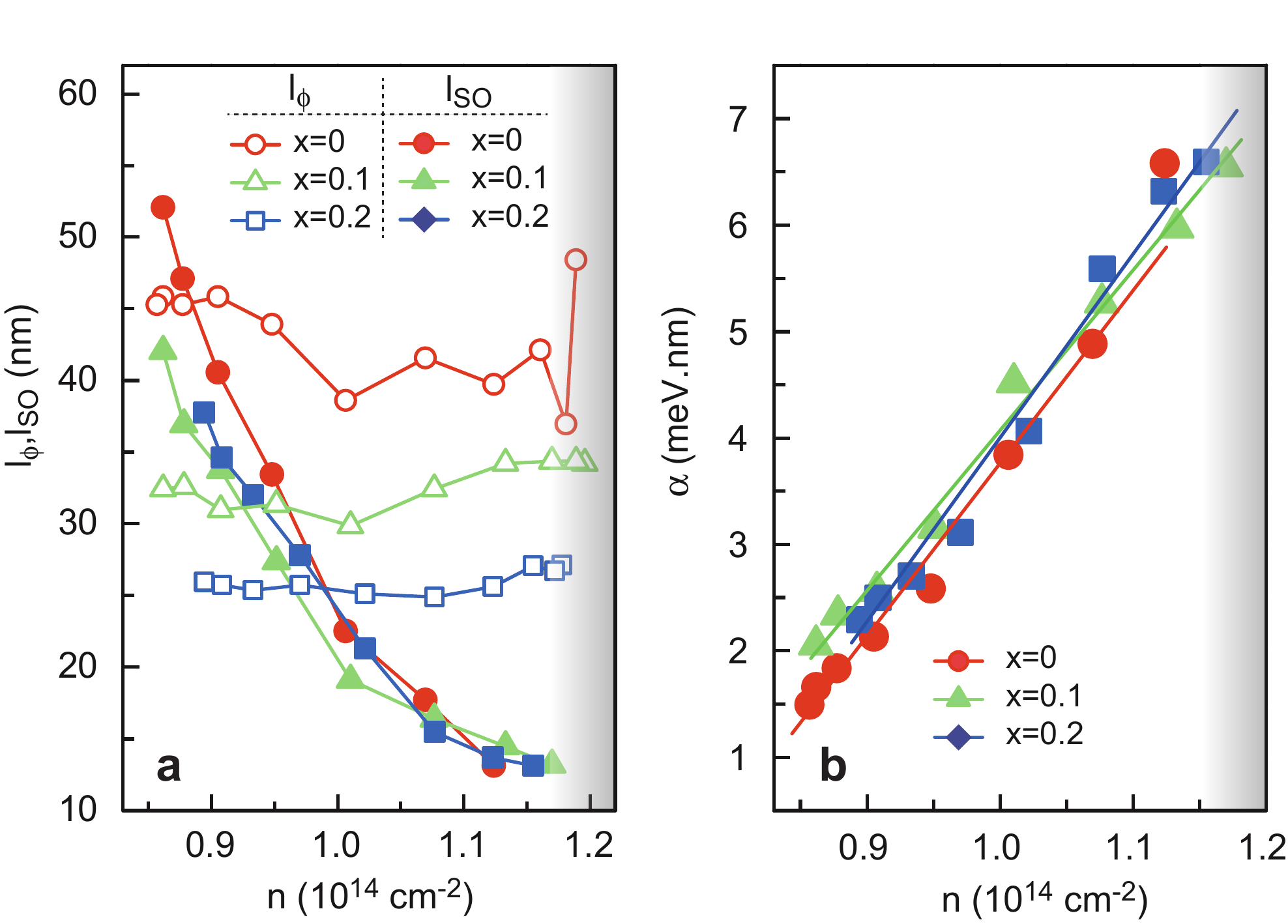}
\end{center}
\vskip -0.7cm
\caption{\label{fig1} (a) Spin-orbit precession length $\ell_\mathrm{SO}$  (filled symbols) and phase coherence length $\ell_\phi$ (open symbols) as a function of the total carrier density $n$. Beyond carrier density 1.16x10$^{14}$ $e^{-}\cdot\mbox{cm}^{-2}$ the estimated values of  $\ell$$_{SO}$  are not reliable since scattering is dominated by orbital magneto-conductance. (b) Spin-orbit coupling constant $\alpha$ (filled symbols) as a function of $n$ for the three samples. Solid lines are linear fits of the data. In the grey zones in Fig. 4a and 4b, $\ell$$_{SO}$,  $\ell$$_\phi$ and $\alpha$ are determined with reduced accuracy (see text). }
\end{figure}

Since $B_\mathrm{SO}$ and  $\ell$$_{SO}$ do not vary with disorder (i.e. with $\tau_e$), their evolution with gate voltage must be mainly due to the modification of the coupling constant $\alpha$ upon gating. In the case of a Rashba spin-orbit interaction, $\alpha$  can directly be extracted from $B_\mathrm{SO}$ : $\alpha =(e\hbar^{3}B_{SO})^{1/2}/m$\cite{perel}.  Assuming an electron effective mass of 0.7m$_e$, which corresponds to d$_{xy}$ sub-bands mainly populated, we find that $\alpha$ increases linearly with the total carrier density for the three samples (Fig. 4b). Changing the gate voltage modifies the carrier density and self-consistently the shape of the quantum well in which the 2-DEG is confined\cite{biscaras2}. Through Maxwell-Gauss equation, the carrier density of the 2-DEG can be related to the interfacial electric field  by $n\simeq\frac{\epsilon}{e}E_z-n_t$ where $\epsilon$ is the dielectric constant of \STO at the interface and $n_t$ is the carrier density of non-mobile charges trapped in the \STO substrate. Our results show that for the three samples, the coupling constant $\alpha$ increases linearly with $E_z$ in agreement with a Rashba spin-orbit interaction. The slope is similar to the one already reported in top-gate experiments \cite{hurand}.\\
 
In summary, we have investigated the effect the substitution of Al by Cr on superconductivity and spin orbit coupling in LaAl$_{1-x}$Cr$_x$O$_3$/SrTiO$_3$ interfaces for three samples corresponding to $x$=0, 0.1 and 0.2. The main effect of Cr doping is to induce disorder in the interfacial quantum well which leads to a decrease of the electronic elastic scattering time of the 2-DEG,  without significant modification of the carrier density.  A suppression of superconductivity is observed by increasing the Cr doping consistent with a Finkelstein\textquoteright s reduction of T$_c$ induced by disorder and electron-electron interactions. By analyzing the magnetoconductance of the 2-DEG as a function of magnetic field, we showed that  the spin diffusion length $\ell_\mathrm{SO}=\sqrt{D\tau_\mathrm{SO}}$ is essentially independent of the disorder (i.e. ${\tau_e}$) corresponding to a D'Yakonov-Perel mechanism of spin relaxation ($\tau_{SO}\sim\frac{1}{\tau_e}$). In addition, we found that the spin-orbit coupling constant $\alpha$ increases linearly with the interfacial electric field $E_z$ which is controlled by the gate voltage, as expected for a Rashba  spin-orbit interaction.

\paragraph*{Acknowledgements}

The authors gratefully thank M. Grilli and S. Caprara for stimulating discussions.This work has been supported by the R\'egion Ile-de-France in the framework of CNano IdF, OXYMORE and Sesame programs, by Delegation G\'enerale \`a l'Armement (S.H. PhD Grant), by CNRS through a PICS program (S2S) and ANR JCJC (Nano-SO2DEG). Part of this work has been supported by the IFCPAR French-Indian program (contract 4704-A). Research in India was funded by the CSIR and DST, Government of India. 
\indent

\clearpage
\large{\textbf{Supplementary Material}}\\

\normalsize
\textbf{Part I :  Variation of carrier density extracted from capacitance measurement. }\\

\LAO/\STO interfaces display multiple type of carrier transport which affects the linearity of the Hall voltage with magnetic field at strong electrostatic doping \cite{kim,ohtsuka}. To extract reliably the carrier density in the 2-DEG, we performed Hall effect measurements as a function of $V_\mathrm{G}$ combined with gate capacitance measurements \cite{biscaras2}. For negative gate voltages, the Hall voltage is linear with magnetic field (see inset Figure 1b in the main text) and the carrier density is correctly extracted from the slope of V$_\mathrm{Hall}$  (ie n$_\mathrm{Hall}$= IB/eV$_\mathrm{Hall}$ where $I$ is the bias current  and $B$ the magnetic field). The variation of carrier density as a function of $V_\mathrm{G}$  can then be retrieved from the integral of the capacitance $C(V_\mathrm{G})$ between the gate and the 2-DEG :
\begin{equation}
n (V_\mathrm{G})=n_\mathrm{Hall}(V_\mathrm{G}=-200V)+\frac{1}{eA}\int_{-200}^{V_\mathrm{G}}C(V_\mathrm{G})dV
\end{equation}
where $A$ is the area of the capacitor (ie the area of the sample). As seen in the Figure 1b (main text), $n$ matches n$_\mathrm{Hall}$ for negative $V_\mathrm{G}$ and extrapolate the curve for positive gate voltages.\\

\begin{figure}[htb]
\includegraphics[width=8.5cm]{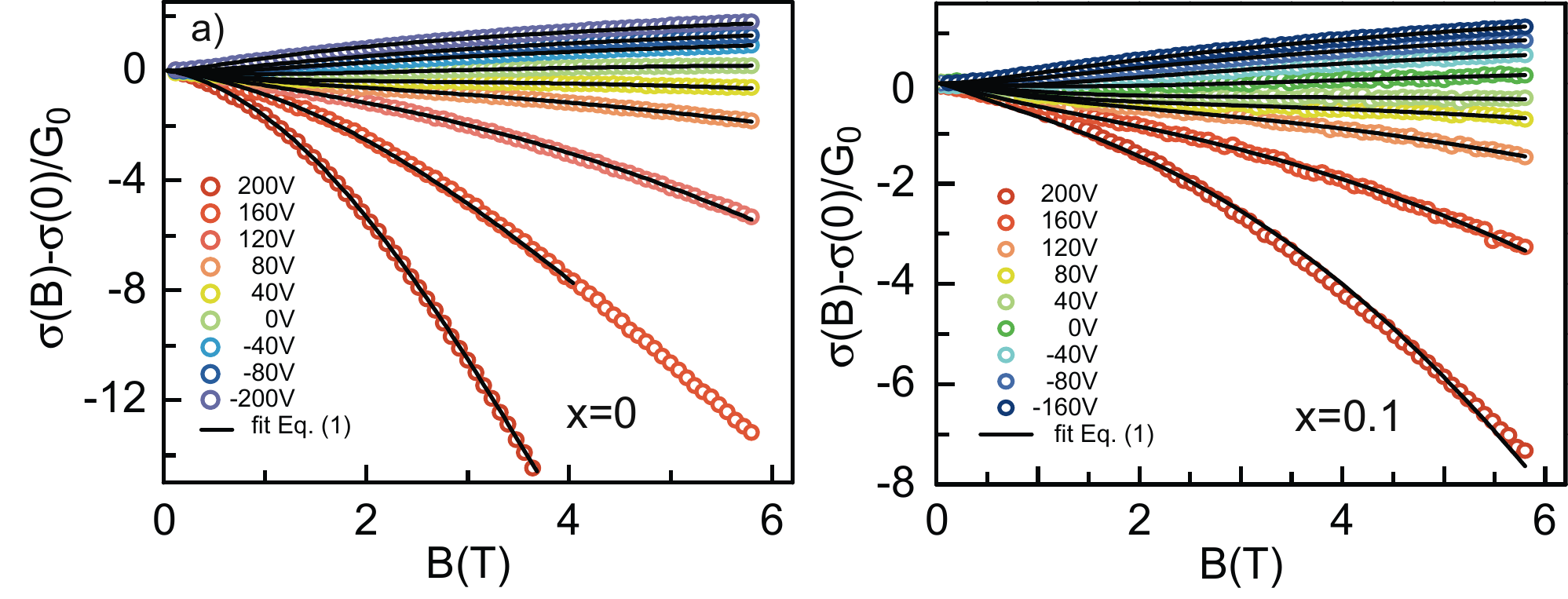}
\caption{Normalized magneto-conductance as a function of out-of-plane magnetic field $B$ of the $x$=0 (panel a) and $x$=0.1 (panel b) samples for different gate voltages $V_\mathrm{G}$. Symbols correspond to data and solid lines to with Eq. (1). }
\label{FigureS1}
\end{figure}

\textbf{Part II :  Reduction of $T_c$ by disorder : Finkelstein's fit}\\

According to Finkelstein\textquoteright s theory \cite{finkelstein}, disorder generates an enhancement of electron-electron interaction that lowers $T_c$. This later follows a universal law that depends on the sheet resistance and the parameter $\gamma=\ln\frac{1}{T_{c0}\tau}$.
\begin{eqnarray}
\lefteqn{T_{c}=}\nonumber\\
& &T_{c0}\exp[\gamma +(1/\sqrt{2r})\ln [\large(1/\gamma +r/4-\sqrt{r/2})/\nonumber\\
& & (1/\gamma +r/4+\sqrt{r/2})]]
\end{eqnarray}
where $r=Re^{2}/(\pi h)$. The best fit shown in Figure 2d of the main text is obtained for a fitting parameter  $\gamma$ = 10.5.\\

\textbf{Part III :  Magneto-conductance of the 2-DEG}\\

The magneto-conductance $\sigma(B)$ of the 2-DEG  was measured as a function of a magnetic field applied perpendicular to sample plane for the three samples. The experimental curves are fitted with Maekawa-Fukuyama formulae \cite{maekawa,hurand} in the diffusive regime (see Eq. (1) in the main text). The results are presented in Figure 5 for the $x$=0 and $x$=0.1 samples (magneto-conductance of sample $x$=0.2 is shown in the Figure 3 of the main text.\\

\textbf{Part IV :  Variation of the $B_{\phi}$ parameter as a function of $V_\mathrm{G}$. }\\

 Figure 3c of the main text, shows that the inelastic field $B_\phi$ does not show any significant dependance with $V_\mathrm{G}$ for the three samples, but increases with doping $x$. This is a direct consequence of enhanced disorder, and therefore reduced elastic scattering time $\tau_e$. Indeed, B$_\phi$ is inversely proportional to $\tau_e$ through the diffusion constant $D$. If the inelastic scattering time $\tau_\Phi$ is independent of $x$ (since it is dominated by electron-electron interactions \cite{hurand,biscaras}) and since the carrier density does not change with $x$, then we expect 
 B$_{\phi}(x)$/B$_{\phi}(x=0)$=$\tau_{e}(x=0)/\tau_{e}(x)$=$R(x)/R(x=0)$. This is rather well verified in Figure 6 within experimental errors, especially in the underdoped region of the phase diagram where single band transport occurs.

\begin{figure}[htb]
\includegraphics[width=8cm]{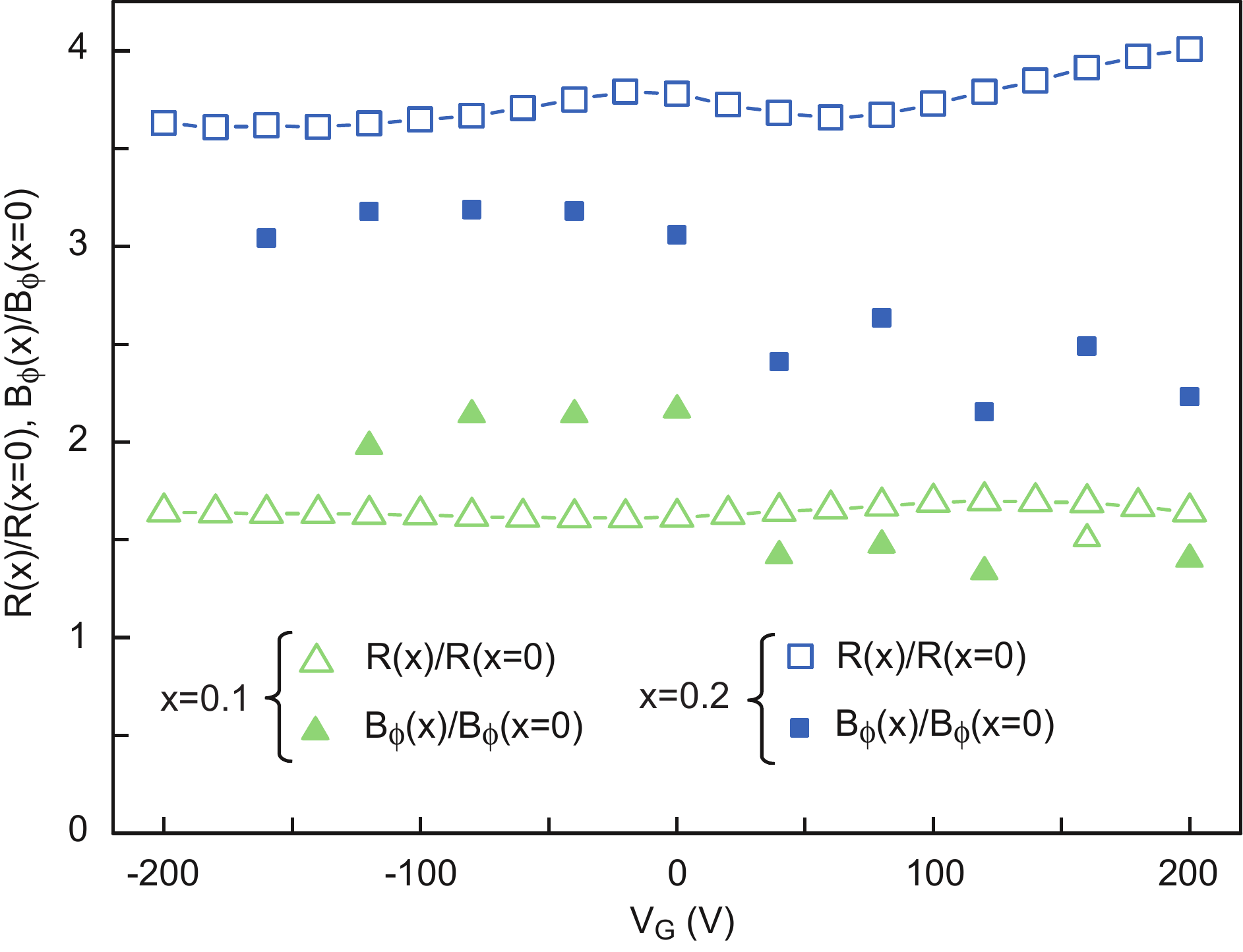}
\caption{Sheet resistance of the $x=0.1$ and $x=0.2$ samples normalized by the sheet resistance of the $x=0$ sample, $R(x)/R(x=0)$, plotted  as a function of $V_\mathrm{G}$ (open symbols) and  $B_{\phi}$ parameter of the $x=0.1$ and $x=0.2$ samples  normalized by $B_{\phi}$ of the $x=0$ sample, $B_{\phi}(x)/B_{\phi}(x=0)$, plotted as a function of $V_\mathrm{G}$ (solid symbols).}
\label{FigureS2}
\end{figure}

\end{document}